\documentclass[aps,twocolumn,prl,superscriptaddress]{revtex4-1}

\usepackage{epsfig}
\usepackage{amsfonts}
\usepackage{amssymb}
\usepackage{mathrsfs}
\usepackage{theorem}
\usepackage{amsmath}
\usepackage{times}

\usepackage{ifpdf}
\ifpdf
\usepackage{epstopdf}   
\fi

\newtheorem{theorem}{Theorem}

\newcommand{\ket}[1]{\ensuremath{\vert#1\rangle}}

\newcommand{\kb}[2]{\ensuremath{\vert #1 \rangle \langle #2 \vert}}

\renewcommand{\vec}[1]{\ensuremath{\mathbf{#1}}}

\newcommand{\cliff}{\ensuremath{\sim_c}}

\def\id{\mbox{\small 1} \!\! \mbox{1}}

\def\id{{\mathchoice {\rm 1\mskip-4mu l} {\rm 1\mskip-4mu l} {\rm 1\mskip-4.5mu l} {\rm 1\mskip-5mu l}}}

\begin{document}
\title{Unifying gate-synthesis and magic state distillation}

\author{Earl T.\ Campbell}
\author{Mark Howard}
\affiliation{Department of Physics \& Astronomy, University of Sheffield, Sheffield, S3 7RH, United Kingdom.}
\email{earltcampbell@gmail.com}

\begin{abstract}
The leading paradigm for performing computation on quantum memories can be encapsulated as distill-then-synthesize.  Initially, one performs several rounds of distillation to create high-fidelity magic states that provide one good $T$ gate, an essential quantum logic gate.  Subsequently, gate synthesis intersperses many $T$ gates with Clifford gates to realise a desired circuit.  We introduce a unified framework that implements one round of distillation and multi-qubit gate synthesis in a single step. Typically, our method uses the same number of $T$-gates as conventional synthesis, but with the added benefit of quadratic error suppression. Because of this, one less round of magic state distillation needs to be performed, leading to significant resource savings.

\end{abstract}


\maketitle  

Development of quantum computers has intensified, spurred on by the prospect that fully fault-tolerant devices are within reach.   A major impetus has been new theoretical advances showing practical designs of fault-tolerant devices can tolerate up to one percent noise~\cite{dennis02}. The topological surface code or toric code is the most widely known breakthrough, which allows for a robust storage of quantum information.  Augmenting the surface code from a static memory to a computer requires additional information processing gadgets. Fault-tolerant information processing can be achieved by a two-step process.  In the first step, logical qubits are distilled from noisy resources into high-fidelity magic states~\cite{BraKit05}.  Each magic state can provide a fault-tolerant $T$-gate, also known as a $\pi/8$ phase gate.  In the second step, gate-synthesis techniques decompose any desired unitary into a sequence of many $T$-gates interspersed with Clifford gates.  This approach to processing quantum information can be paraphrased as distill-then-synthesize.  Most leading laboratories are following designs~\cite{herrera10,Fowler12,Nickerson14} within this paradigm of distill-then-synthesize combined with surface codes.  While alternative ideas to magic state distillation exist~\cite{bombin:topoDistil,bombin:codeDeform,Paetznick13,bombin15}, so far they lack the appealing high tolerance to noise~\cite{brown15,bravyi15}.  We propose a framework where both distillation and synthesis occur simultaneously, which we call synthillation.
	
\begin{figure}[t]
	\includegraphics{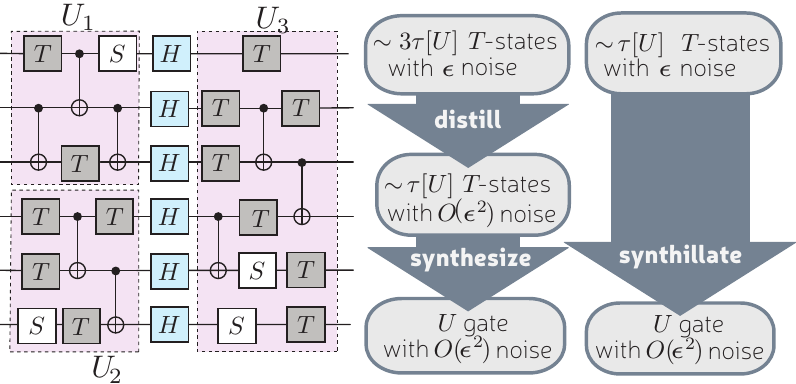}
	\caption{(a) Toy example circuit $U$ divided into subcircuits $\{ U_1, U_2, U_3 \}$ and interspersed by Hadamard gates.  Subcircuits contain only control-NOT, $S$ and $T$ gates. (b) Schematic explaining the 1/3 resource saving of synthillation over distill-then-synthesize (using BHMSD). The $T$ cost of synthesizing $U$ using \cite{amy16} is denoted $\tau[U_F]$. }
\label{fig1}
\end{figure}

Fault-tolerance protocols come with a price-tag, an overhead of extra qubits.  Consequently, genuinely useful applications may need millions or billions of physical qubits.  Improved protocols for magic state distillation~\cite{Meier13,Bravyi12,jones13b} and gate synthesis~\cite{amy13,amy14,amy16,maslov16,kliuchnikov13,RS14,bocharov15} have reduced resource overheads, but the cost remains formidable and further overhead reduction is extremely valuable. Notable is the Bravyi-Haah magic state distillation (BHMSD)  protocol \cite{Bravyi12} that converts $3k+8$ magic states into $k$ magic states with quadratic error suppression.  For large computations, with between $10^{10}$ and $10^{15}$ logical operations, the required precision can be reached by concatenating BHMSD two or three times, assuming an initial physical error rate of order $\sim 0.1\%$.  Multilevel distillation is an effective tool when many rounds are required~\cite{jones13b}.  Gate synthesis has advanced on two fronts.  For synthesis of single qubit gates, optimal protocols have been found~\cite{kliuchnikov13,RS14,bocharov15}.  For multi-qubit circuits generated by CNOT and $T$ gates, optimal and exact synthesis results exist~\cite{amy13,amy14,amy16,maslov16}.   This multiqubit gate set requires Hadamards to acquire universality, and so gate-synthesis can be applied to subcircuits separated by Hadamards as shown in Fig.~(\ref{fig1}a).  Progress on distill-then-synthesize schemes has principally been achieved by refining the two component processes separately.  However, there exists schemes for directly distilling more exotic resources thereby obviating the need for subsequent synthesis. In Refs.~\cite{duclos12,duclos15,campbell16}, resource states for small-angle single-qubit rotations are distilled, whereas in Refs.~\cite{jones13b,eastin13,Paetznick13} the resource state for a Toffoli gate is distilled.  While inspirational to our approach, these techniques do not apply to a general class of multi-qubit circuits and are formally quite distinct from any gate-synthesis protocols.

Here we present a general framework for implementing error-suppressed multiqubit circuits generated by CNOT and $T$ gates.  Our technique fuses notions of phase polynomials used in multiqubit exact synthesis~\cite{amy13,amy14,amy16} with Bravyi and Haah's triorthogonal matrices~\cite{Bravyi12} into a single unified framework. This sets it apart from previous alternatives~\cite{duclos12,duclos15,jones13b,eastin13,campbell16} to the distill-then-synthesize paradigm, which share little formalism in common with gate-synthesis methods.  Our approach also yields practical benefits;  in the worst case using synthillation is never more expensive than conventional distill-then-synthesize but, for a broad and important class of circuits, synthillation effectively eliminates the need for one round of distillation. Measuring resource costs by noisy $T$-states consumed, our approach can reduce magic state factories by greater than a factor of 3. A full architecture specific resource analysis, also counting all Clifford operations, is beyond our current scope but could reveal much greater resource savings.
 

The group of gates producible from CNOT and $T$ gates can always~\cite{amy13,amy14,amy16} be decomposed into a CNOT circuit followed by a diagonal unitary
\begin{equation}
\label{Eq_Uf}
    U_F = \sum_{\vec{x} \in \mathbb{Z}_2^k} \omega^{F(\vec{x})} \kb{\vec{x}}{\vec{x}},
\end{equation}
where $\ket{\vec{x}}$ are basis states labeled by binary strings $\vec{x}^T=(x_1,x_2,\ldots, x_k)$, we use $\omega=e^{i \frac{\pi}{4}}$ throughout and $F$ is a polynomial $F:\mathbb{Z}_2^k \rightarrow \mathbb{Z}_8$ of a particular weighted form
\begin{align}    
\label{function_form}
    F(\vec{x}) & = L(\vec{x}) + 2 Q(\vec{x}) + 4 C(\vec{x}), 
\end{align}
where  and $L, Q$ and $C$ are linear, quadratic and cubic polynomials respectively. For example, a unitary with a single $T$ gate, controlled-$S$ gate (where $S=T^2$) and control-control-Z (CCZ) gate is described by the polynomial $x_2+2x_1x_2+4x_1x_3x_4$.  These unitaries form a group that we label as $\mathcal{D}_3$ since they are the diagonal gates from the $3^{\mathrm{rd}}$ level of the Clifford hierarchy~\cite{CliffHier}.
We find a special role is played by the CCZ gate, which differs by Cliffords from the Toffoli and corresponds to a cubic monomial $4x_1 x_2 x_3$.  Doubled functions $2F$ correspond to $U_{2F}$ that are diagonal Clifford gates~\cite{amy14,amy16}.  Therefore, a unitary $U_{F}$ is always Clifford equivalent to $U_{F + 2 \tilde{F}}$ for any $\tilde{F}$ of the above form, and we denote this Clifford equivalence relation as $F \cliff F + 2 \tilde{F}$.  We denote $\tau[U_F]$ as the ancilla-free $T$-count for exact synthesis of $U_F$.   We also define $\mu[U_F]$ to be the minimum $\tau[V]$ over all decompositions of $U_F=VW$ where $W$ is composed purely of CCZ gates.  This is enough to state our main result.
\begin{theorem}
\label{thm_prots}
Let $\{U_1, U_2,\ldots U_l \}$ be a set of diagonal unitaries in the family $\mathcal{D}_3$, and $U_F := \otimes U_j$.  The synthillation protocol can implement $\{U_1, U_2,\ldots U_l \}$ with probability $1-n\epsilon+ O(\epsilon^2)$ and error rate $O(\epsilon^2)$ using 
\begin{equation}
    n = \tau[ U_F ] + 2 \mu[ U_F ] + \Delta \leq 3 \tau[U_F] + \Delta, 
\end{equation}
noisy $T$-states of initial error rate $\epsilon$ where $0 \leq \Delta \leq 11$.
\end{theorem}
The constant $\Delta$ is bounded and so negligible in the large circuit limit. The $\epsilon$ quantifies imperfection of magic states, and not synthesis precision since this is an exact synthesis problem. The expected resource cost is $n / p_{\mathrm{suc}}$, which approaches $n$ for small $\epsilon$.  Regarding the quantity $\mu[U_F]$, we have $\mu[U_F] \leq \tau[U_F] $ by setting $V=U$, which leads to $n \lesssim 3\tau[ U_F ]$.   Therefore, our approach is never more expensive than using a round of BHMSD followed by gate synthesis, which uses $\sim 3 \tau[ U_F ]$ resources.  

\begin{figure}[t]
	\includegraphics{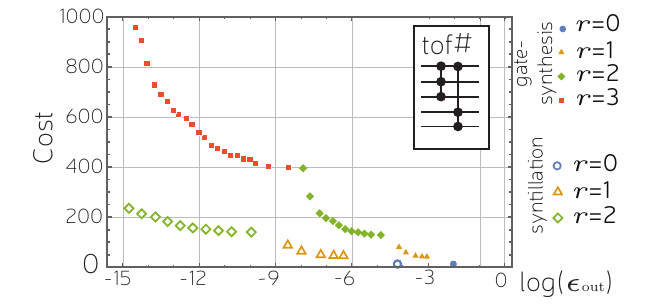}
	\caption{Average number of raw magic $T$ states, with initial error $0.1\%$, required to produce a $\mathrm{tof}^{\#}$ gate with final error rate $\epsilon_{\mathrm{out}}$.  We compare $r$ rounds of BHMSD followed by gate-synthesis (filled shapes) with $r$ rounds of BHMSD followed by synthillation (empty shapes). This particular 5-qubit gate has a $T$-count $\tau[U_F]=11$ and synthillation reduces costs by a factor $\sim 3.6$ across a broad range of target error rates.  Inset shows $\mathrm{tof}^{\#}$ decomposed as 2 CCZ gates.}
\label{fig2}
\end{figure}

Synthillation offers roughly a one-third saving over distill-then-synthesize whenever $\mu[ U_F ] \ll \tau[ U_F ]$  (see  Fig.~\ref{fig1}b for a schematic comparison).  This maximum saving can be attained when the circuit consists of CCZ gates as we can then choose $W=U_F$ and $V=\id$, entailing $\mu[U_F]=0$.  Resource assessments are slightly adjusted when $\epsilon$ is non-negligible, but this typically amplifies the merit of synthillation.  As an example, Fig.~\ref{fig2} shows the exact resource cost of implementing the $\mathrm{tof}^{\#}$ gate with polynomial $4x_1(x_2 x_3+ x_4 x_5)$. Gates of this form -- using only Toffoli gates, CNOT gates and NOT gates -- appear frequently in Shor's algorithm and many other quantum algorithms (subcircuits for implementing the necessary reversible logic and quantum arithmetic appear in e.g., \cite{gossett98,Draper:2006,Abdessaied:2016,bocharov16}). This is an explicit class of circuits where synthillation has a significant advantage because $0=\mu[U_F] \ll \tau[U_F]$ and so $n \ll 3 \tau[U_F]$.  
More generally,  $\tau[ U_F ]$ may grow quadratically with the number of qubits~\cite{amy16}, whereas $\mu[U_F]$ can grow at most linearly~\cite{Camp16c}.  Therefore, there is a large class of complex circuits where $\mu[U_F] \ll \tau[U_F]$, and so again synthillation offers a free round of error suppression. 

Synthillation proceeds by fault-tolerantly preparing the state $\ket{\psi_F}=U_F \ket{+}^{\otimes k}$.  Since $U_F$ is in $\mathcal{D}_3$, the resource $\ket{\psi_F}$ can be used to deterministically teleport the gate $U_F$ into a quantum computation~\cite{CliffHier,GC01a}.  When $U_F$ is broken into components $\otimes_j U_j$ each can be teleported to any required location in the computation.  We begin by defining a class of quantum codes and some concise notation. Let $G$ be a full rank binary matrix with $n$ columns and $k+s$ rows that is partitioned into sub-matrices $K$ and $S$, which we denote as $G=(\frac{K}{S})$.  From this matrix, we define a quantum code with logical basis states
\begin{align}
    \ket{\vec{x}_L} &:= \frac{1}{2^{s/2}}  \sum_{\vec{y} \in \mathbb{Z}_2^{s}}  \ket{K^T\vec{x} \oplus S^T\vec{y}}, \label{eqn:logical}\\
       \left( K^T\vec{x} \oplus S^T\vec{y} \right)_j&:=\sum_{i=1}^{k}K_{i,j}x_i+\sum_{i=1}^{s}S_{i,j}y_i \bmod 2 \nonumber
\end{align}
This is an $[[n,k,d]]$ code where $n$ is the number of columns in $G$, $k$ is the number of rows in $K$, and $d$ is the distance.  We can always pad $G$ with extra rows to get a square invertible matrix $J$, and given such a matrix there exists~\cite{Dehaene03,patel03,maslov07} a CNOT circuit realising $\ket{\vec{z}}\rightarrow \ket{J^T \vec{z}}$.  We call any such circuit an encoder $E_G$ since $E_G \ket{\vec{x},\vec{y},\vec{0}}=\ket{K^T\vec{x} \oplus S^T\vec{y}}$ and so $E_G \ket{\vec{x}}\ket{+}^{\otimes s} \ket{\vec{0}}=\ket{\vec{x}_L}$. We require quantum codes with logical operators of a peculiar nature.  We say a code is $F$-quasitransversal if there exists a diagonal Clifford $C$ such that $C T^{\otimes n}$ acts as a logical $U_F$ on the codespace i.e., $CT^{\otimes n} \ket{\vec{x}_L}=\omega^{F(\vec{x})}\ket{\vec{x}_L}$.  The code must be tailored to the target unitary, just as circuit synthesis depends on the target unitary.  We can quickly establish a sufficient condition on $G$ so that $F$-quasitransversality holds. 
First note that for all $\vec{e}\in \mathbb{Z}_2^n$ we have $T^{\otimes n}\ket{\vec{e}}=\omega^{|\vec{e}|}\ket{\vec{e}}$ where $|\vec{e}|:=\sum_{i=1}^ne_i$. We combine this observation with Eq.~\eqref{eqn:logical} to find
\begin{align}
T^{\otimes n} \ket{\vec{x}_L}&
=\frac{1}{2^{s/2}}\sum_{\vec{y}}\omega^{|K^T\vec{x} \oplus S^T\vec{y}|}\ket{K^T\vec{x} \oplus S^T\vec{y}}.
\end{align}
Note that any diagonal Clifford $\tilde{C}$ acts as 
\begin{equation}
\label{eqn:Ctilde}
	\tilde{C} \ket{\vec{x}}\ket{\vec{y}}\ket{\vec{0}}^{\otimes n-k-s}=\omega^{2\tilde{F}(\vec{x},\vec{y})}\ket{\vec{x}}\ket{\vec{y}}\ket{\vec{0}}^{\otimes n-k-s} ,
\end{equation}
for some $\tilde{F}$.  Defining the Clifford $C:=E_G \tilde{C} E_G^\dagger$, we have 
\begin{align}
C T^{\otimes n} \ket{\vec{x}_L}&
= \frac{1}{2^{s/2}} \sum_{\vec{y}}\omega^{|K^T\vec{x} \oplus S^T\vec{y}|+2\tilde{F}(\vec{x},\vec{y})}\ket{K^T\vec{x} \oplus S^T\vec{y}} \nonumber\\
&=\frac{1}{2^{s/2}} \sum_{\vec{y}}\omega^{F(\vec{x})}\ket{K^T\vec{x} \oplus S^T\vec{y}}=\omega^{F(\vec{x})}\ket{\vec{x}_L} , \nonumber
\end{align}
where the last line holds provided there exists an $\tilde{F}$ so that
\begin{align}
\omega^{F(\vec{x},\vec{y})}=\omega^{|K^T\vec{x} \oplus S^T\vec{y}|+2 \tilde{F}(\vec{x})} \quad \forall (\vec{x},\vec{y}). \label{eqn:condition}
\end{align}
or, in other words, provided $|K^T\vec{x} \oplus S^T\vec{y}| \cliff F(\vec{x})$.
%
%
%
%
%
We later return to providing explicit constructions of $G$.  

\begin{figure}[t]
	\includegraphics{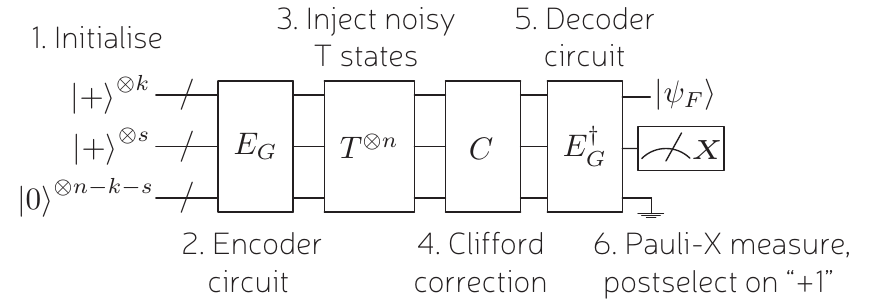}
 	\caption{Synthillation preparation of $\ket{\psi_F}$ magic state.  The Clifford correction is $C=E_G \tilde{C} E_G^\dagger$ where  $\tilde{C}$ is defined by Eq.~\eqref{eqn:Ctilde}.}	
\label{Prot_distil}
\end{figure}

Given a $F$-quasitransversal quantum code, the first stage of synthillation is to use it to prepare the multi-qubit state $\ket{\psi_F}=U_F \ket{+}^{\otimes k}$ using the protocol described in Fig.~\ref{Prot_distil}. In the absence of noise, preparation of $\ket{\psi_F}$ follows immediately from $F$-quasitransversality.  We consider the effect of  $T$ gates suffering Pauli-$Z$ noise, which can be assumed due to standard twirling arguments.  To describe $Z$ operators acting on many qubits we use $Z[\vec{e}]:=\otimes_{j=1}^n Z_j^{e_j}$ where $\vec{e}$ is some binary vector.  Therefore, at step 3 we must add the operator $Z[\vec{e}]$ with probability $p(\vec{e})=\epsilon^{|\vec{e}|}(1-\epsilon)^{n-|\vec{e}|}$.  For a given $Z[\vec{e}]$ and definition of encoder unitaries, it follows that
\begin{equation}
	E_G^\dagger	Z[\vec{e}] E_G = Z[ K \vec{e}] \otimes  Z[ S \vec{e}]  \otimes Z[ M \vec{e}] .
\end{equation}
The matrix $M$ corresponds to row padding used to make $G$ a square matrix. The component $Z[ M \vec{e}]$ will soon vanish so we do not dwell its exact form. Using that  $Z$ operators commute with the diagonal Clifford $C$, we find
\begin{align}
	 & C E_G^\dagger Z[\vec{e}] T^{\otimes n} E_G  \ket{+}^{\otimes k+s}\ket{\vec{0}} \\ \nonumber &=  (Z[K \vec{e}]U_F \ket{+}^{\otimes k})(Z[S \vec{e}] \ket{+}^{\otimes s}) \ket{\vec{0}},
\end{align}
where we have used $Z\ket{\vec{0}}=\ket{\vec{0}}$ to eliminate $Z[ M \vec{e}]$.   In step 6, we measure the qubits in the state $Z[S \vec{e}] \ket{+}^{\otimes s}$ declaring the SUCCESS outcome only if $S \vec{e}=(0,0,\ldots0)^T$.  Therefore, the success probability is
\begin{equation}
	p_{\mathrm{suc}} = \sum_{\vec{e} :  S \vec{e}=(0,\ldots 0)^T} \epsilon^{|\vec{e}|}(1-\epsilon)^{n-|\vec{e}|}.
\end{equation}
When successful, the output state is $Z[K \vec{e}]U_{F}\ket{+}^{\otimes k}$ which is the correct state whenever $K\vec{e}=(0,0,\ldots )^T$.  Therefore, the normalised error rate is
\begin{equation}
	\epsilon_{\mathrm{out}} =1-\frac{1}{p_{\mathrm{suc}}} \sum_{\vec{e} :  K \vec{e}=(0,\ldots 0)^T} \epsilon^{|\vec{e}|}(1-\epsilon)^{n-|\vec{e}|}.
\end{equation}
For a distance $d$ code, we have that if $S\vec{e}=(0,0,\ldots )^T$ and $K \vec{e} \neq (0,0,\ldots )^T$ then $|\vec{e}| \geq d$. This allows us to conclude the scaling $\epsilon_{\mathrm{out}} = O(\epsilon^d)  $.  

We have established a fault-tolerant process for preparing $U_F\ket{+}^{\otimes k}$, assuming a nontrivial $F$-quasitransversal code.  The second major ingredient in our proof is the notion of phase polynomials from the gate-synthesis literature~\cite{amy13,amy14,amy16}, which we now review.   Phase polynomials are used to rewrite functions $F(\vec{x})$ from  Eq.~\eqref{function_form}
\begin{equation}
F(\vec{x})\rightarrow P_\textbf{a}(\vec{x}) = \sum_{\vec{u} \in \mathbb{Z}_2^{r} } a_{\vec{u}} [\bigoplus x_j u_j \pmod{2}] \pmod{8} ,
\end{equation}
where we index the integer elements of vector $\textbf{a}$ with the label $\vec{u} \in \mathbb{Z}_2^{r}$. For example, a suitable expansion for CCZ is
\begin{align}
4x_1x_2x_3&\rightarrow x_1+x_2+x_3+(x_1\oplus x_2\oplus x_3) \label{eqn:CCZ_decomp}\\
&\quad +7(x_1\oplus x_2)+7(x_2\oplus x_3)+7(x_1\oplus x_3). \nonumber
\end{align}
It is known~\cite{amy13,amy14,amy16} that for every weighted polynomial of the form in Eq.~\eqref{function_form}, there exists a $P_\textbf{a}$ such that $P_\textbf{a}(\vec{x})=F(\vec{x})$ for all $\vec{x}$.  Conversely, every phase polynomial equals some weighted polynomial.  The values of $\textbf{a}$ are only important modulo 2 because of the Clifford equivalence $P_\textbf{a} \cliff P_{[\textbf{a} \pmod 2]}$.  Once we have a phase polynomial $P_\vec{a}$, one can construct a gate-synthesis circuit using a quantity of $T$-gates equal to $|\vec{a} \pmod{2}|=\sum_{\vec{u}}[a_{\vec{u}} \pmod{2} ]$. For example, the expansion in Eq.~\eqref{eqn:CCZ_decomp} shows that CCZ can be synthesized with seven $T$ gates, seen by counting the number of terms with odd coefficients. A phase polynomial representation of a function $F$ is not always unique,  so we minimise over all $\textbf{a}$ such that $P_\textbf{a}=F$. Amy and Mosca~\cite{amy16} showed that this optimisation problem is equivalent to decoding a Reed-Muller code and gives the optimal $T$-count attainable using ancilla-free gate synthesis over the gate set $\{ $CNOT$, T, S \}$.      

A key insight here is that we can relate phase polynomials with matrices arising from quantum codes.  Defining $A$ to be any $k$-by-$n$ binary matrix where the column vector $\vec{u}$ appears once if $a_{\vec{u}}=1 \pmod{2}$, one can quickly verify that 
\begin{align}
|A^T \vec{x}|= P_{[\textbf{a} \pmod 2]}(\vec{x}) \cliff P_{\textbf{a}}(\vec{x}). \label{eqn:AmatPhase}
\end{align}
 Setting $G=A$ we can construct a trivial quantum code with $F$-quasitransversality, and this provides an explicit method of implementing $U_F$ using $T$-gates.  As such, we call $A$ a gate-synthesis matrix for $U_F$. The number of qubits in the code equals the number of columns in $A$, which equals the number of odd-valued components, $a_\vec{u}$, in the vector $\vec{a}$.   If $U_F$ can be synthesised with $\tau[U_F]$ gates then there is $P_\vec{a}$ enabling us to construct an optimal $A$ with $\tau[U_F]$ columns. This presents a fresh perspective on gate-synthesis.

We now finalise the proof of our main result by providing explicit $G$ matrices.   Our constructions depend on several features of the unitary, and we begin with the case where $U_F$ is a CCZ circuit so that $F$ is a homogeneous cubic polynomial.  Let $A$ be the optimal gate-synthesis matrix for $U_F$, which we momentarily assume has an even number of columns, then  
\[ 
G = \left( \begin{array}{c}
K   \\ \hline
S \end{array} \right) = \left( \begin{array}{c}
A   \\ \hline
\rule[-.3\baselineskip]{0pt}{11pt}  \vec{1}^T \end{array} \right), \quad \vec{1}^T=(1,1,\ldots,1)
\]
generates an $F$-quasitransversal distance 2 code using $n=\tau[U_F]$ qubits.  The first step in the proof is to note
\begin{align}
|K^T\vec{x} \oplus S^T \vec{y}|& = | A^T \vec{x} \oplus (y_1\vec{1})| ,\\ \nonumber
	&= | A^T \vec{x}| +  |(y_1 \vec{1})| - 2 y_1 |A^T\vec{x}|,
\end{align}
where we have used $\alpha \oplus \beta = \alpha + \beta - 2 \alpha \beta$.  From Eq.~\eqref{eqn:AmatPhase} we know $|A^T \vec{x}| \cliff  F(\vec{x})$. Therefore, we need the remaining terms to be Clifford.  Since $|y_1  \vec{1}| = \tau[U_F] y_1$ and $\tau[U_F]$ is assumed even, this term is Clifford.  For the third term we again use $|A^T \vec{x}| \cliff  F(\vec{x})$ so that 
\begin{align}
	2 y_1 |A^T\vec{x}| = 2 y_1 F(\vec{x}) + 2 y_1 (2\tilde{F}(\vec{x}))
\end{align}
We already know $2\tilde{F}$ is Clifford and multiplying it by $2y_1$ preserves Cliffordness since the degree of terms increases by 1, but the coefficient is doubled.  For the term $2 y_1 F(\vec{x})$ we use that $F$ is homogeneous cubic, and Eq.~\eqref{function_form} required that cubic terms carry a prefactor of 4, combined with the prefactor $2y_1$ we find this term vanishes modulo 8.  This proves $F$-quasitransversality.  We assumed that $\tau[U_F]$ is even, because our argument used that $|\vec{1}|$ is even.  We can deal with odd $\tau[U_F]$ by padding $A$ with a column of zeros and using the above, leading to a small additive cost $n=\tau[U_F]+1$.  The proof is almost identical.  

We now turn to more general $U_F$, and introduce $U_F= VW$ where $W$ is a CCZ circuit.  Again, $A$ is the gate-synthesis matrix for $U_F$, but we now also use $B$ as the gate-synthesis matrix for $V$.  We define the $G$ matrix
\begin{equation}
	G = \left( \begin{array}{c}
K   \\ \hline
S \end{array} \right)
= \left(  \begin{array}{ccccccccccc}
A & B & B & \vec{c} & \vec{c} & \vec{c} & \vec{c} & 0 & 0 & 0 & 0 \\  \hline
1 & 1 & 0 & 1 & 0 & 0 & 1 & 1 & 0 & 0 & 1  \\
1 & 0 & 1 & 0 & 1 & 0 & 1 & 0 & 1 & 0 & 1 \\
0 & 0 & 0 & 1 & 1 & 1 & 1 & 1 & 1 & 1 & 1 \\
\end{array}\right), 
\end{equation}
where $\vec{c}$ is fixed so that $\sum_j c_j x_j$ equals the linear terms in $F$.  It follows that provided $\tau[U_F]$ and $\mu[U_F]$ are even, the quantum code associated with $G$ is $F$-quasitransversal using $n=\tau[U_F] + 2\mu[U_F] + 8$ qubits.  To prove this we must show $|K^T \vec{x} \oplus S^T \vec{y}| \cliff F(\vec{x}) \pmod 8$.  Though a more complex $G$ is needed for more generic functions, and the proof is necessarily longer, the proof technique is the same in essence. One again converts from modular to standard arithmetic and removes Clifford terms until only $F(\vec{x})$ remains.  In the proof we use that $\tau[U_F]$ and $\mu[U_F]$ are even, but all  cases can be handled with  slight variants of the above $G$ matrix.


We have focused on synthillation processes where the input resources are $T$ gates, and the output is a very different object, a general unitary in the $\mathcal{D}_3$ family.  However, this general technique includes when the output are also $T$ gates.  It is informative to reflect on how synthillation relates to triorthogonal matrices used in BHMSD.  When $U_F=T^{\otimes k}$, with even $k$, we have that $A$ and $B$ are the identity matrix and $\vec{c}$ is the all ones column vector.  This gives, up to column permutation, the same $G$ matrix employed by Bravyi and Haah.  We see our $G$ matrices are generalizations of triorthogonal matrices. In a longer paper~\cite{Camp16c}, we give a more extensive discussion of synthillation and several additional results.  Notably, we show the optimal $U_F=VW$ decomposition can be efficiently solved, which leads to several interesting insights into optimal gate-synthesis including an efficient algorithm for finding near-optimal circuit decompositions.  We also show that $\tau$ is not always additive, with a single CCZ gate requiring 7 $T$-gates but $N$ such gates need only $6N+1$ $T$-gates, and so synthillation uses $6N+2$ $T$-states.  

This work shows the possibility of significant resource savings by considering distillation and synthesis in a more holistic manner.  This resource reduction is additional to savings from optimised gate-synthesis~\cite{amy13,amy14,amy16} and module checking~\cite{Ogorman16}. We quantified resources by $T$-states consumed, which is a common approximation, with a full resource count~\cite{RHG01a,fowler13,Ogorman16,maslov16} being the natural next step.  We remark that the formalism can be extended to higher levels of the Clifford hierarchy, but we found this yielded no significant benefits. 

\textit{Acknowledgements.-} This work was supported by the EPSRC (grant EP/M024261/1).  We thank Ben Brown, Joe O'Gorman, Matthew Amy, and Dmitri Maslov for comments on the manuscript.


%

\end{document}